\documentclass[12pt]{article}
\usepackage{a4wide,amssymb,cite}
\pdfoutput=1

\usepackage{a4wide,amssymb,graphicx}
\usepackage{epsfig}
\usepackage[usenames,dvipsnames]{color}
\usepackage{slashed}
\usepackage{amssymb,cite,graphicx}
\usepackage{slashed}
\usepackage{amsmath,bm,bbm}
\usepackage{amsfonts}
\usepackage[titletoc,title]{appendix}
\usepackage[small]{caption}
\usepackage[margin=1in]{geometry}
\usepackage[multiple]{footmisc}
\usepackage{mathtools}
\usepackage{slashed}
\usepackage[nottoc]{tocbibind}
\usepackage{xcolor}

\usepackage{tikz}
\usetikzlibrary{shapes,arrows,positioning,automata,backgrounds,calc,er,patterns}
\usepackage[compat=1.1.0]{tikz-feynman}
\usetikzlibrary{trees}
\usetikzlibrary{decorations.pathmorphing}
\usetikzlibrary{decorations.markings}

\newcommand{\be}{\begin{equation}}
\newcommand{\ee}{\end{equation}}
\newcommand{\bea}{\begin{eqnarray}}
\newcommand{\eea}{\end{eqnarray}}

\def\circa#1{\,\raise.3ex\hbox{$#1$\kern-.75em\lower1ex\hbox{$\sim$}}\,}

\begin{document}

\begin{titlepage}
%
%


%

\begin{centering}
\vspace{1cm}
{\Large {\bf Pseudo-Nambu-Goldstone inflation \vspace{0.2cm}\\ with twin waterfalls}} \\

\vspace{1.5cm}

{\bf Hyun Min Lee$^\dagger$ and Adriana Menkara$^\sharp$ }
\\
\vspace{.5cm}

{\it  Department of Physics, Chung-Ang University, Seoul 06974, Korea.}

\vspace{.5cm}


\end{centering}
\vspace{2cm}

\begin{abstract}
\noindent
We propose a pseudo-Nambu-Goldstone inflation where twin waterfall fields coupled to the inflaton are responsible for the graceful exit from inflation by a waterfall transition. In this scenario, the $Z_2$ symmetry for the waterfall fields and the inflaton protects  the inflaton potential against dangerous quantum corrections coming from the waterfall couplings.  We show that there is a wide range of natural model parameters and initial conditions for a successful inflation and discuss the reheating process by the perturbative decay of the waterfall field with a $Z_2$ invariant Higgs portal. We also present a microscopic model for the inflaton couplings to the twin waterfall fields in the context of a dark QCD with light and heavy quarks.

\end{abstract}

\vspace{3cm}

\begin{flushleft} 
$^\dagger$Email: hminlee@cau.ac.kr \\
$^\sharp$Email: amenkara@cau.ac.kr 
\end{flushleft}

\end{titlepage}

\section{Introduction}

Cosmic inflation solves various problems in the Standard Big Bang cosmology such as the  horizon problem, homogeneity, isotropy problems, etc. Quantum fluctuations during inflation leave footprints in the isotropies of Cosmic Microwave Background (CMB) as measured precisely by Planck \cite{planck} and seed the large-scale structure in cosmology. Thus, inflationary model building and pursuing its cosmological signatures have been an active research field in particle physics and cosmology in the last decades.  

Natural inflation \cite{natural} has drawn a particular attention among a plethora of inflation models in the literature. In this model, the inflaton is identified as a pseudo-Nambu-Goldstone boson (pNGB) or an axion-like scalar field, which stems from a spontaneously broken global symmetry as in the case for pions in QCD. Thus, the shift symmetry makes quantum corrections to the inflaton potential under control at the perturbative level, and the perturbative effects such as QCD-like instantons can generate a desirable inflaton potential in the low energy. However, the slow-roll inflation and the correct CMB normalization require the scale of the spontaneous breakdown of the global symmetry (or the axion decay constant) to be far beyond the Planck scale, so the effective field theory for inflaton could break down due to unknown quantum gravity effects \cite{qgravity,sundrum0}. Furthermore, according to the weak gravity conjecture \cite{wgc},  gravity is supposed to be the weakest force, ruling out a trans-Planckian axion decay constant in quantum gravity \cite{wgc-axion}. Therefore, there have been some extensions of natural inflation where the effective axion decay constant for inflation becomes trans-Planckian due to the alignment for multiple axions \cite{multiaxion}. 

We propose a model for natural inflation, which consists of a pNGB inflaton and two waterfall fields. In this scenario, a slow-roll inflation is driven by the pNGB field while the waterfall fields are decoupled. But, in the presence of the inflaton-dependent masses for the waterfall fields, inflation ends due to the tachyonic instability of the waterfall fields being developed during inflation \cite{hybrid}.  Thus, in this type of hybrid natural inflation, there is no need of a trans-Planckian  axion decay constant for natural inflation. 
Instead, we need to make sure that the necessary couplings of the waterfall fields to the inflaton do not generate dangerous quantum corrections to the inflaton potential.  To this purpose, we introduce a $Z_2$ mirror symmetry for the waterfall fields such that  the inflaton potential is insensitive to the quantum corrections coming from the couplings between the inflaton and the heavy waterfall fields. Such a discrete symmetry can be protected in quantum gravity \cite{discrete} so that the consistency of the inflaton couplings  is ensured for the hybrid natural inflation. 
 
Introducing the general waterfall sector respecting the shift symmetry for the inflaton and the $Z_2$ symmetry simultaneously, we discuss the dependence of the inflationary predictions on the inflaton parameters and the waterfall couplings. In this model, we also study the vacuum structure and the reheating process from the perturbative decay of the waterfall field with a Higgs portal coupling as well as the consequence of the reheating period for the inflationary prediction. 
We provide a novel microscopic model for realizing the pNGB inflation with twin waterfall fields in a QCD-like theory and show the non-decoupled effects of heavy dark quarks for the waterfall couplings.
There is one appendix dealing with the general vacuum structure for the waterfall fields.

\section{The model}

We consider a pseudo-Nambu-Goldstone boson $\phi$ as the inflaton and two real scalar fields $\chi_1, \chi_2$ as waterfall fields in the hybrid inflation scenarios.

We first introduce the scalar potential for the inflaton in the following,
\bea
V(\phi,\chi_1,\chi_2) =V_I(\phi)+V_W(\phi,\chi_1,\chi_2)
\eea
where the inflaton potential is given by
\bea
V_I(\phi) &=& V_0 + \Lambda^4 \cos\Big(\frac{\phi}{f} \Big), \label{inflaton} 
\eea
and $V_W(\phi,\chi_1,\chi_2)$ is the waterfall field part which is model-dependent.
For the hybrid inflation, we choose  $V_0\gtrsim \Lambda^4$, so we need another sector (with waterfall fields) for the graceful exit.

Imposing a $Z_2$ discrete symmetry \cite{sundrum} with
\bea
\phi\longrightarrow-\phi,  \qquad  \chi_1\longleftrightarrow \chi_2,
\eea
we also take the waterfall part of the scalar potential for the hybrid inflation with the inflaton couplings in the following form,
\bea
V_W(\phi,\chi_1,\chi_2)&=&-\frac{1}{2}\mu^2\sin\Big(\frac{\phi}{2f} \Big) (\chi^2_1-\chi^2_2) +\frac{1}{2} m^2_\chi (\chi^2_1+\chi^2_2) -
\alpha^2 \chi_1\chi_2\nonumber \\
&&+\frac{1}{4} \lambda_\chi (\chi^4_1+\chi^4_2) + \frac{1}{2} {\bar\lambda}_\chi \chi^2_1 \chi^2_2.  \label{full}
\eea
Here, we note that there could be more cubic and quartic terms for the waterfall fields, respecting the $Z_2$ symmetry and affecting the vacuum structure. But, those terms do not affect the waterfall transition, so our inflationary prediction remains valid.
In a later section, we will present the details on the microscopic description of the waterfall field couplings to the inflaton in the context of a dark QCD. Similar hybrid inflation models with a real scalar field were also considered in the context of the brane inflation \cite{hmlee} and the pNGB inflation \cite{sundrum,jeong}. 

During inflation, there is no VEV for the waterfall fields during inflation, but there is a mass mixing between the waterfall fields for $\alpha\neq 0$.
Then, we identify the inflaton-dependent mass eigenvalues for the waterfall fields  by
\bea
m^2_1(\phi) &=& m^2_\chi - \sqrt{ \mu^4 \sin^2\Big(\frac{\phi}{2f} \Big)+\alpha^4},  \label{chi1mass}\\
m^2_2(\phi)&=& m^2_\chi +\sqrt{ \mu^2 \sin^2\Big(\frac{\phi}{2f} \Big)+\alpha^4}, \label{chi2mass}
\eea
and the mixing angle $\theta$ between the waterfall fields depends on the inflaton field by
\bea
\sin2\theta(\phi)=\frac{2\alpha^2}{m^2_2(\phi)-m^2_1(\phi)}.
\eea
Here, we can keep the kinetic terms for the waterfall fields in the approximately canonical forms during the slow-roll inflation.
For $\alpha=0$, there is no mixing between the waterfall fields, so we can just keep track of the waterfall field $\chi_1$ to determine the end of inflation. 

For $\phi<\phi_c$ where $\phi_c=2f\arcsin(\sqrt{m^4_\chi-\alpha^4}/\mu^2)$ with $\sqrt{m^4_\chi-\alpha^4}<\mu^2$ and $\alpha<m_\chi$, the slow-roll inflation takes place.
In this case, the waterfall fields are heavy enough for $m_\chi>H_I$ with $H_I$ being the Hubble scale during inflation, so we can describe the slow-roll inflation by the inflaton potential given in eq.~(\ref{inflaton}).
At $\phi=\phi_c$ the waterfall field with mass $m_1$ starts becoming unstable, ending the inflation even if the slow-roll condition for the inflaton direction is not violated.  

Due to the couplings of the waterfall fields to the inflaton, the inflaton potential receives loop corrections. The one-loop Coleman-Weinberg potential for the inflaton is given in cutoff regularization with cutoff scale $M_*$, as follows,
\bea
V_{\rm CW} &=&\frac{1}{64\pi^2}\sum_{i=1,2} \left[2m^2_{\chi_i}M^2_* -m^4_{\chi_i} \ln\bigg( \frac{e^{\frac{1}{2}}M^2_* }{m^2_{\chi_i}}\bigg)  \right] \nonumber \\
&\simeq&\frac{1}{16\pi^2}\,m^2_\chi M^2_* -\frac{1}{64\pi^2}\bigg[m^4_\chi +\mu^4\sin^2\Big(\frac{\phi}{2f}\Big)+\alpha^4\bigg] \ln  \frac{M^2_*}{m^2_\chi}.  \label{CW}
\eea 
Then, the constant vacuum energy proportional to $M^2_*$ must be renormalized to get the desirable inflation energy. On the other hand, the quadratically divergent part of the inflaton potential is cancelled between the waterfall fields due to the $Z_2$ discrete symmetry, and the logarithmically divergent terms of the inflaton potential can be ignored during inflation as far as $\mu^2\lesssim 8\pi \Lambda^2$ is satisfied.  

Consequently,  as far as the mass parameters in the waterfall field sector satisfies $H^2_I\lesssim \mu^2\sim m^2_\chi\lesssim 8\pi \Lambda^2$, the waterfall fields remain decoupled and affect the inflaton mass little.
Moreover, for a QCD-like phase transition with  $\Lambda\ll f$ and the inflaton mass,  $\sqrt{|m^2_\phi|}= \frac{\Lambda^2}{f} $,  we maintain a natural hierarchy of scales in our model,
\bea
\sqrt{|m^2_\phi|} \ll H_I\ll \mu\sim m_\chi\ll \sqrt{8\pi} \Lambda \ll f.  
\eea 
Then, as we vary the inflation scale, we can also scale the other mass parameters freely subject to the above hierarchy to get a consistent natural inflation of hybrid type. 
We also remark that the bound on the waterfall fields couplings is consistent with the condition that a QCD-like phase transition occurs during inflation, that is, $\Lambda>H_I$.

\section{Inflationary predictions}

We now discuss the inflationary predictions of the pNGB inflation with twin waterfall fields.
Ignoring the classical dynamics of the waterfall fields during inflation, we focus on the slow-roll inflation and the condition for the waterfall transition.  

First, from the inflaton potential in eq.~(\ref{inflaton}), the slow-roll parameters are given by
\bea
\epsilon &=& \frac{M^2_P \Lambda^8 \sin^2(\phi/f)}{2f^2 (V_0+\Lambda^4\cos(\phi/f))^2}, \\
\eta &=& -\frac{M^2_P \Lambda^4 \cos(\phi/f)}{f^2 (V_0+\Lambda^4\cos(\phi/f))}.
\eea
The number of efoldings is also obtained as
\bea
N&=&\frac{1}{M_P} \int^{\phi_c}_{\phi_*} \frac{1}{\sqrt{2\epsilon}}\, d\phi  \nonumber \\
&=& \frac{f^2}{2M^2_P \Lambda^4}\, \bigg[V_0 \ln \Big(\tan^2\Big(\frac{\phi_c}{2f}\Big)\Big)+\Lambda^4\ln \Big(\sin^2\Big(\frac{\phi_c}{f}\Big)\Big) \bigg] -(\phi_c\to \phi_*) 
\eea
where $\phi_*,\phi_c$ are the inflaton field values at the horizon exit and at the end of inflation, respectively.

Now we consider the case with $V_0\gg \Lambda^4$ for which the couplings between the inflaton and the waterfall fields are necessary to end inflation. 
In this case, the slow-roll parameters and the number of efoldings are approximated to
\bea
\eta_* &\simeq& -\frac{M^2_P \Lambda^4}{f^2V_0 }\, \cos(\phi_*/f), \\
\epsilon_* &\simeq & \frac{M^2_P \Lambda^8}{2f^2V^2_0}\,  \sin^2(\phi_*/f), \\
N &\simeq&  \frac{f^2 V_0}{M^2_P \Lambda^4}\,\ln \Big(\frac{\tan(\phi_c/(2f))}{\tan(\phi_*/(2f))} \Big). \label{Nefold}
\eea
As a result, the spectral index and the tensor-to-scalar ratio can be determined by
\bea
n_s &=&1+2\eta_*-6\epsilon_*,  \label{ns}\\
r&=&16\epsilon_*. \label{r}
\eea
The CMB normalization, $A_s=\frac{1}{24\pi^2} \frac{V_0+\Lambda^4}{\epsilon_* M^4_P}\simeq 2.1\times 10^{-9}$, leads to
\bea
r=3.2\times 10^7\,\cdot\frac{V_0}{M^4_P}.  \label{cmb}
\eea
In order to get the spectral index consistent with Planck data, $n_s=0.967\pm 0.0037 $ \cite{planck}, we need to choose $2\eta_*\simeq -0.0033$ because  $ \epsilon_*\ll |\eta_*|$ in our case. The critical value $\phi_c$ of the inflaton should satisfy $ \phi_*\lesssim \phi_c\lesssim f$ for the number of efoldings $N=50-60$ to solve the horizon problem. 
Moreover, the Planck bound on the tensor-to-scalar ratio, $r<0.036$ \cite{keck}, gives rise to the upper bound on $H_I<4.6 \times 10^{13}\,{\rm GeV}$.

In order to check the parameter space for inflation in our model, from $H^2_I\simeq V_0/(3M^2_P)$ and eqs.~(\ref{ns}), (\ref{Nefold}) and (\ref{cmb}), it is more convenient to choose the following parametrization,
\bea
n_s&\simeq& 1+2\eta_*\simeq 1- \frac{|m^2_\phi|}{3H^2}\, \cos(\phi_*/f), \quad |m^2_\phi|= \frac{\Lambda^4}{f^2}, \\
N&=& \frac{\cos(\phi_*/f)}{|\eta_*|}\,  \ln \Big(\frac{\tan(\phi_c/(2f))}{\tan(\phi_*/(2f))} \Big), \\
\frac{H_I}{f}&=&2.9\times 10^{-4}\,\Big|\eta_*  \tan(\phi_*/f)\Big|, \label{cmb2}
\eea
together with the condition determining the end of inflation,
\bea
\sin\Big(\frac{\phi_c}{2f}\Big) = \frac{\sqrt{m^4_\chi-\alpha^4}}{\mu^2}. 
\eea
From eq.~(\ref{cmb2}), we find that the axion decay constant and the Hubble scalar during inflation is correlated.
Taking $|\eta_*|=0.033/2$ to get the consistent spectral index and $\cos(\phi_*/f)=0.95$, we get $f\simeq 6.4\times 10^5 H_I$, which is much larger than the Hubble scale, so the global symmetry responsible for the PNG inflaton is broken during inflation.

\begin{figure}[t]
\centering
\includegraphics[width=0.40\textwidth,clip]{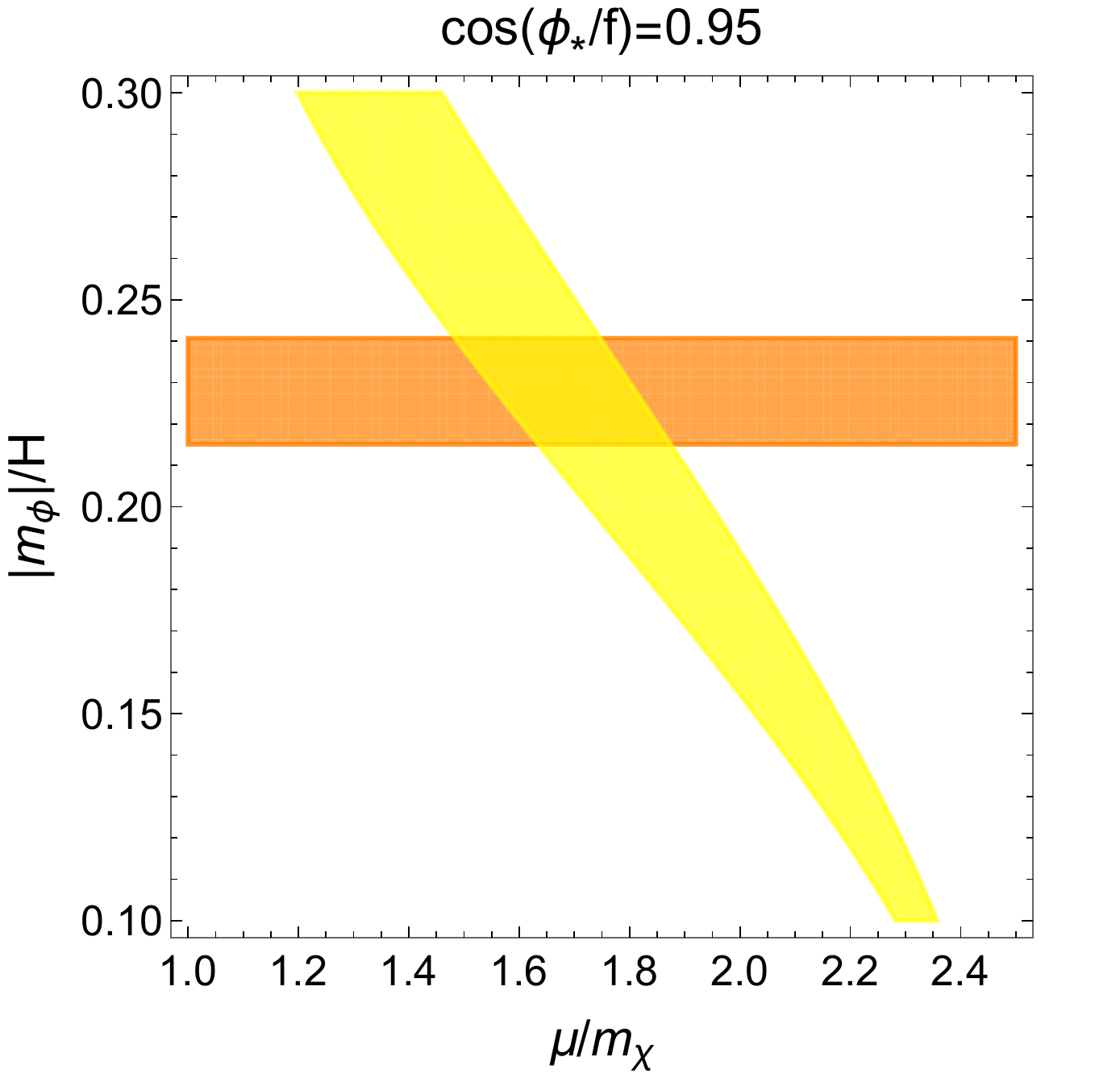}\,\,\,\,\,\,
\includegraphics[width=0.40\textwidth,clip]{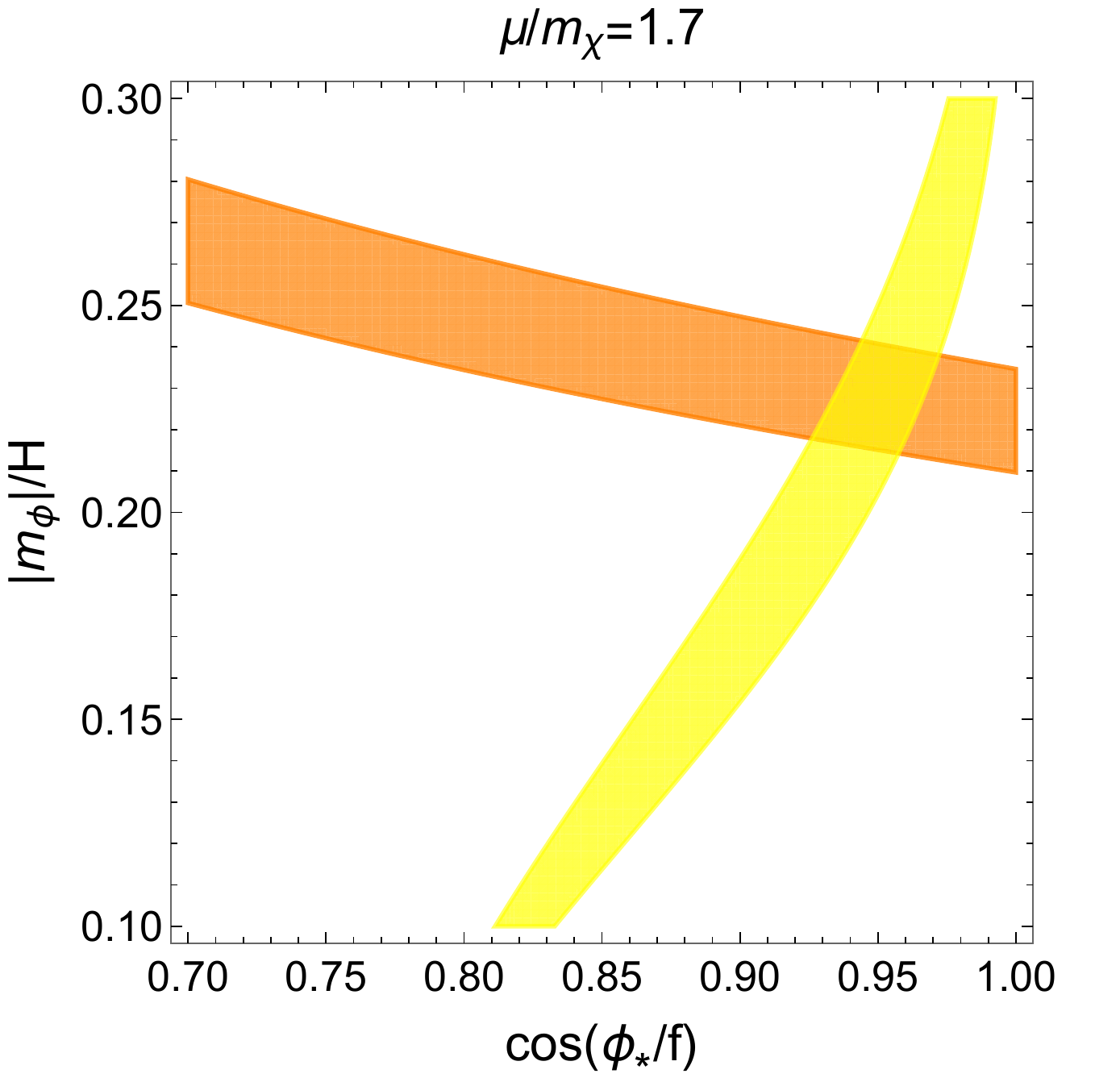}
\caption{Parameter space for inflation in $|m_\phi|/H$ vs $\mu/m_\chi$ on left and $|m_\phi|/H$ vs $\cos(\theta_*/f)$ on right. The orange region is consistent with Planck data within $1\sigma$ and  the number of efoldings is given by $N=40-60$ in the yellow region. Successful inflation is possible in the overlapping region.  We chose  $\alpha=0$, $\cos(\theta_*/f)=0.95$ on left and $\mu/m_\chi=1.7$ on right.  
}
\label{fig:inf1}
\end{figure}

In Fig.~\ref{fig:inf1}, we show the parameter space for the successful inflation in our model, for $|m_\phi|/H_I$ vs $\mu/m_\chi$ in the left plot and $|m_\phi|/H_I$ vs $\cos(\theta_*/f)$ in the right plot.  The orange region is consistent with Planck data within $1\sigma$ and the yellow region indicates the number of efoldings between $40$ and $60$. Thus, the overlapping region between the two regions leads to a successful inflation. We took $\cos(\theta_*/f)=0.95$ on left and $\mu/m_\chi=1.7$ on right, and the CMB normalization was taken into account.  Although we chose $\alpha=0$ in Fig.~\ref{fig:inf1}, we only have to replace $m^2_\chi$  by $\sqrt{m^4_\chi-\alpha^4}$ for a nonzero $\alpha$. 
As we increase $\mu/m_\chi$, the waterfall transition would take place at a smaller $\phi_c$, so the inflation would have started closer to the origin in order to satisfy the correct number of efoldings in eq.~(\ref{Nefold}).

\begin{figure}[t]
\centering
\includegraphics[width=0.45\textwidth,clip]{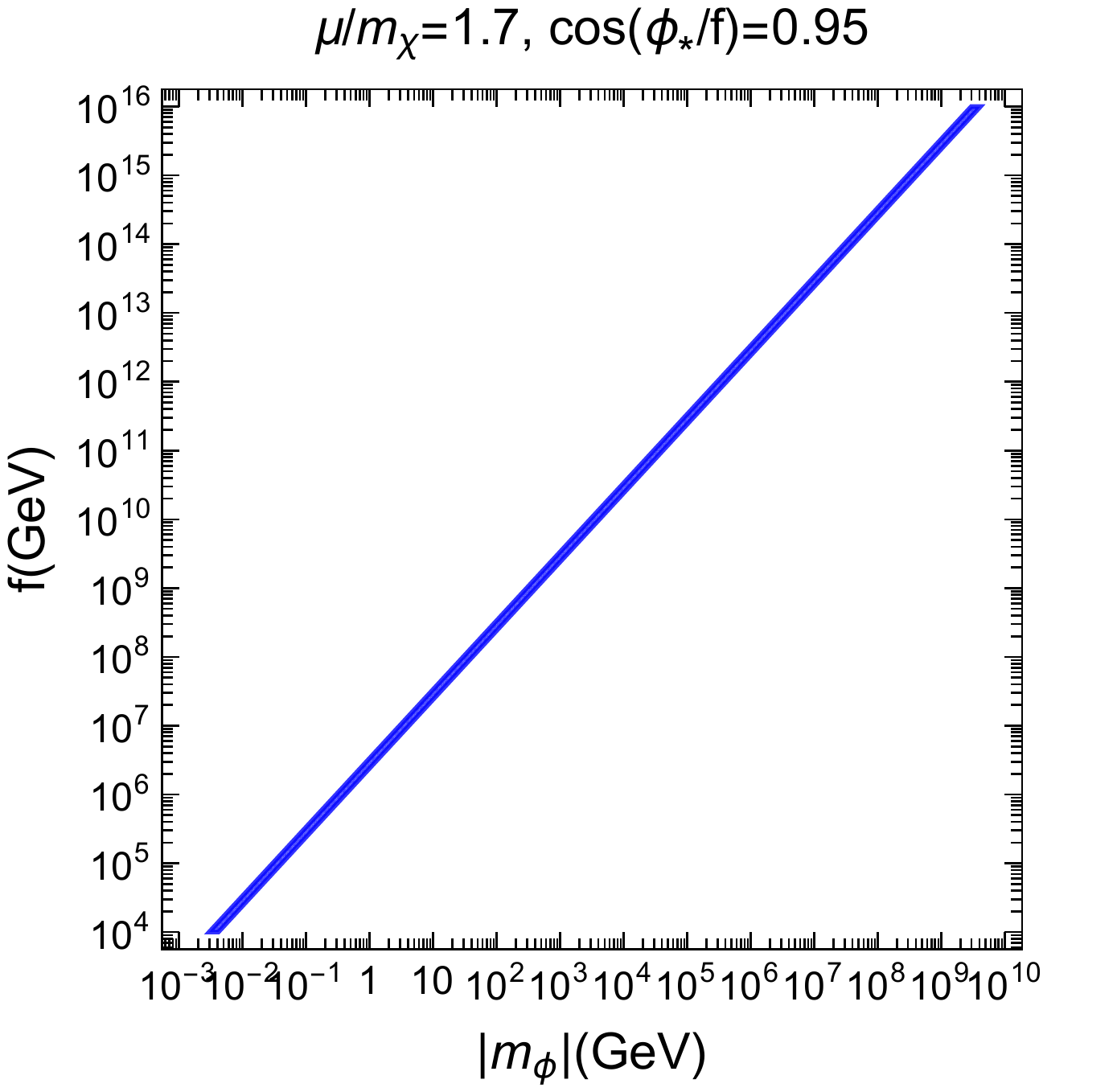}
\caption{Parameter space for inflation in $f$ vs $|m_\phi|$. The blue line is consistent with Planck data within $1\sigma$ and  the number of efoldings, $N=40-60$.   We chose  $\alpha=0$, $\mu/m_\chi=1.7$ and $\cos(\theta_*/f)=0.95$. 
}
\label{fig:inf2}
\end{figure}

In Fig.~\ref{fig:inf2}, we present the parameter space for the axion decay constant $f$ vs $|m_\phi|$ in blue line, which is consistent with Planck data within $1\sigma$. We took  $\mu/m_\chi=1.7$ and $\cos(\theta_*/f)=0.95$, but other choices close to them lead to a similar correlation. Thus, we find that for $f=10^4\,{\rm GeV}-10^{16}\,{\rm GeV}$, the successful inflation is achieved for the values of $|m_\phi|=\Lambda^2/f$ between $3.5\times 10^{-3}\,{\rm GeV}$ and $3.5\times 10^9\,{\rm GeV}$, which correspond to the range for the QCD-like condensation scale, $5.9\,{\rm GeV}\lesssim \Lambda\lesssim 5.9\times 10^{12}\,{\rm GeV}$.

For the benchmark point in  Fig.~\ref{fig:inf2}, the Hubble scale $H_I$ varies in the range between $0.016\,{\rm GeV}$ and $1.6\times 10^{10}\,{\rm GeV}$, which leads to a tiny tensor-to-scalar ratio, $4.1\times 10^{-33}\lesssim r\lesssim 4.1 \times 10^{-9}$, from eq.~(\ref{cmb}) with $V_0\simeq 3M^2_P H^2_I$. Moreover, the inflation energy is given by $2.6\times 10^8\,{\rm GeV}\lesssim V^{1/4}_0\lesssim 2.6 \times 10^{14}\,{\rm GeV}$, which is much larger than the QCD-like condensation scale $\Lambda$. We can maintain the inflationary predictions against the quantum corrections coming from the waterfall field coupling, provided that $\mu\ll \sqrt{8\pi} \Lambda=30\,{\rm GeV}-3.0\times 10^{13}\,{\rm GeV}$.
Consequently, there is a large range of the natural parameter space for the successful inflation, satisfying $H_I\ll \mu\ll \sqrt{8\pi} \Lambda$ in our model.  Thus, depending on the input parameters such as the axion decay constant, the waterfall fields with mass of order $\mu$ can be light.

\section{Waterfall transition and reheating}

After the end of inflation, reheating take place due to the decay or scattering of the waterfall fields. 
In this section, we discuss the vacuum structure for the inflaton and the waterfall fields and the reheating dynamics in our model. 

From the scalar potential, we first identify the effective inflaton mass squared from $\frac{\partial^2 V}{\partial\phi^2}$, as follows,
\bea
m^2_{\phi, {\rm eff}} =  -\frac{\Lambda^4}{f^2} \cos\Big(\frac{\langle\phi\rangle}{f}\Big) +\frac{\mu^2}{8f^2}( \langle\chi^2_1\rangle - \langle\chi^2_2\rangle)\sin\Big(\frac{\langle\phi\rangle}{2f}\Big) \label{effinfmass}
\eea
where $\langle\,\,\rangle$ denotes the background field values. We note that the first term  in eq.~(\ref{effinfmass}) corresponds to a tachyonic mass during inflation and the second term in eq.~(\ref{effinfmass})  is an additional contribution to the inflaton mass  after inflation, coming from the waterfall field couplings. 

At the end of inflation, the waterfall field direction with $m^2_1<0$ starts rolling fast at $\phi=\phi_c$, developing a nonzero background field and making the extra contribution to the effective inflaton mass. Thus, the inflaton moves toward a stable minimum near $\phi/f= \pi$, which is the common minimum for the inflaton potential and the waterfall-induced potential.
On the other hand, the waterfall field masses are given by eqs.~(\ref{chi1mass}) and (\ref{chi2mass}) with $\phi>\phi_c$, which are of order $\mu\sim m_\chi$ even after inflation ends.

\subsection{The vacuum structure}

Taking $\alpha=0$ for simplicity \footnote{See Appendix A for the general vacuum structure for $\alpha\neq 0$.}, we find that there is a stable minimum of the potential at $\langle\phi\rangle=v_\phi$,  $\langle\chi_1\rangle=v_1\equiv v_\chi $ and  $\langle\chi_2\rangle=0$, with
\bea
v_\phi &=& \pi f, \\
 v_\chi&=& \sqrt{\frac{\mu^2-m^2_\chi}{\lambda_\chi}}. \label{chivev}
\eea
As a result, we find that  the $Z_2$ symmetry is broken in the vacuum. 

We find that the cosmological constant in the true vacuum  is fine-tuned to the observed value by
\bea
V_{\rm eff}(\chi_1=v_\chi,\chi_2=0) = V_0-\Lambda^4- \frac{1}{4}\lambda_\chi v^4_\chi \simeq 0, \label{zerocc}
\eea
constraining the parameters of the waterfall fields\footnote{If there is a reduction mechanism for the vacuum energy during the waterfall transition, we may make the cosmological constant in the true vacuum to zero. But, in this case, the reheating through the waterfall fields is limited and model-dependent. }.

Next, expanding around the VEV by $\phi=v_\phi+a$  we obtain the inflaton mass as
\bea
m^2_{a} = \frac{1}{f^2} \Big(\Lambda^4+\frac{1}{8} \mu^2 v^2_\chi\Big), \label{inflatonmass} 
\eea
thus the inflaton receives a mass contribution from the VEV of the waterfall field. 
Taking the waterfall fields as $\chi_1=v_\chi+{\tilde\chi}_1$, we also get the mass eigenvalues for the waterfall fields in the vacuum as
\bea
m^2_1&=&  2\lambda_\chi v^2_\chi \nonumber \\
&=&2(\mu^2-m^2_\chi),  \\
m^2_2 &=& \mu^2 + m^2_\chi +{\bar\lambda}_\chi v^2_\chi \nonumber  \\
&=& \mu^2+m^2_\chi +\frac{{\bar\lambda}_\chi }{\lambda_\chi}(\mu^2- m^2_\chi). 
\eea
We also note that the vacuum stability requires $\lambda_\chi>0$ and $\lambda_\chi+{\bar\lambda}_\chi>0$ for ${\bar\lambda}_\chi<0$. 
For $\mu^2/m^2_\chi<3+2({\bar\lambda}_\chi/\lambda_\chi)/(1-{\bar\lambda}_\chi/\lambda_\chi)$, the waterfall field $\chi_2$ is heavier than ${\tilde\chi}_1$. Otherwise, the waterfall field $\chi_2$ is lighter than ${\tilde\chi}_1$. The former case with $m_2>m_1$ is favored for the number of efoldings as discussed in Section 3, so the decay mode of the waterfall field, ${\tilde\chi}_1\to \chi_2\chi_2$, is not open for reheating. 

Moreover, we also identify the leading interaction terms between the inflaton and the mass eigenstates of the waterfall fields, $ {\hat\chi}_{1,2}$, as follows,
\bea
{\cal L}_{\rm int} = \frac{\mu^2}{8f^2}\,v_\chi a^2 {\tilde\chi}_1+ \frac{\mu^2}{16f^2} a^2 ({\tilde\chi}^2_1-{\chi}^2_2)+\cdots. \label{inflaton-int}
\eea
Thus, there is no quadratic divergence in the radiative corrections to the inflaton mass, due to the cancellation between the waterfall fields with the $Z_2$ symmetry.  But, the logarithmic divergence in the radiative corrections to the inflaton mass is present due to the cubic interactions in the first line in eq.~(\ref{inflaton-int}).

We now discuss the interplay of the inflation and the vacuum structure to constrain the parameters of the waterfall fields.
For $V_0\gg \Lambda^4$, eqs.~(\ref{chivev}) and (\ref{zerocc}) give rise to $4V_0/\lambda_\chi\sim v^4_\chi\sim \mu^4/\lambda_\chi^2$. 
Then, the quartic coupling and the VEV for the waterfall fields are related to the dimensionful parameters of the inflation, as follows,
\bea
\lambda_\chi\sim \frac{\mu^4}{4V_0}, \qquad v^2_\chi\sim \frac{4V_0}{\mu^2}. \label{relation}
\eea
As a result, from the condition, $\mu\gtrsim H_I$,  to  keep waterfall fields decoupled during inflation, we have the bounds on the quartic coupling and the VEV for the waterfall fields by $\lambda_\chi\sim \mu^4/(4V_0)\simeq 1.4\times 10^{-20}(\mu/(100H_I))^4 (H_I/10^{5}\,{\rm GeV})^2$ and $v_\chi\sim \sqrt{4V_0/\mu^2}\simeq 0.035 M_P(100H_I/\mu)$. 
Here, we took $f\simeq 6.4\times 10^{5}  H_I$ from  the CMB normalization with $\cos(\phi_*/f)=0.95$.  

Moreover, from eq.~(\ref{inflatonmass}) with $\mu^2 v^2_\chi\sim 4V_0\gg \Lambda^4$ from eq.~(\ref{relation}),  we find that the additional contribution from the waterfall field coupling dominates the inflaton mass as $m^2_a\sim \mu^2 v^2_\chi/(8 f^2)$.

\subsection{Reheating}

After inflation ends, the evolution of scalar fields and the radiation energy density $\rho_R$ is governed by the following set of the Boltzmann equations,
\bea
{\ddot \phi}+3H {\dot \phi}  &=&-\Gamma_\phi {\dot\phi} +\frac{\Lambda^4}{f} \sin\Big(\frac{\phi}{f}\Big)+ \frac{\mu^2}{4f} \cos\Big(\frac{\phi}{2f} \Big) (\chi^2_1-\chi^2_2),  \\
{\ddot\chi}_1 + 3H {\dot\chi}_1  &=&-\Gamma_{\chi_1} {\dot\chi}_1+\mu^2\sin\Big(\frac{\phi}{2f} \Big) \chi_1-m^2_\chi \chi_1-\alpha^2 \chi_2-\lambda_\chi \chi^3_1 -{\bar\lambda}_\chi \chi_1 \chi^2_2, \\
{\ddot\chi}_2 + 3H {\dot\chi}_2 &=&-\Gamma_{\chi_2} {\dot\chi}_2 -\mu^2\sin\Big(\frac{\phi}{2f} \Big) \chi_2-m^2_\chi \chi_2-\alpha^2\chi_1-\lambda_\chi \chi^3_2 -{\bar\lambda}_\chi \chi^2_1 \chi_2, \\
{\dot\rho}_R + 4H \rho_R &=&\Gamma_{\phi} {\dot\phi}^2 +\Gamma_{\chi_1}{\dot\chi}^2_1  +\Gamma_{\chi_2}{\dot\chi}^2_2 ,
\eea
and the Friedmann equation,
\bea
H^2= \frac{\rho_I+\rho_R}{3M^2_P},
\eea
where $\rho_I$ is the sum of energy densities for the inflaton and the waterfall fields, given by 
\bea
\rho_I = \frac{1}{2} {\dot\phi}^2 + \frac{1}{2} {\dot\chi}^2_1 + \frac{1}{2} {\dot\chi}^2_2 +V(\phi,\chi_1,\chi_2).
\eea

Taking $\alpha=0$, we can set $\chi_2=0$ from inflation towards reheating and focus on the dynamics of the inflaton and the waterfall field $\chi_1$.
During reheating, the inflation energy $V_0$ stored in the waterfall field can be transferred to radiation in the presence of the decay of the waterfall field $\chi_1$. On the other hand, since $\Lambda^4\ll V_0$ for the hybrid inflation, we can ignore the reheating from the inflaton field, unless the couplings of the waterfall fields to the visible sector are sufficiently smaller than the one for the inflaton\footnote{We can introduce a linear coupling of the inflaton coupling to gluons by ${\cal L}_{g}=\frac{C}{32\pi^2}\,\frac{\phi}{f} G_{\mu\nu} {\tilde G}^{\mu\nu}$, with $C$ being constant, which is assumed to generate the resulting QCD potential respecting the $Z_2$ symmetry. }. 
Then, we can determine the reheating temperature dominantly from the decay of the waterfall field by
\bea
T_{\rm RH}= \bigg(\frac{90}{\pi^2 g_{\rm RH}} \bigg)^{1/4} \sqrt{M_P \Gamma_{\chi_1}}  \label{RH}
\eea
where $g_{\rm RH}$ is the number of relativistic degrees of freedom at reheating completion and $ \Gamma_{\chi_1}$ is the decay rate for $\chi_1$.

If reheating is not instantaneous, the number of efoldings required to solve the horizon problem \cite{reheating,HiggsR2} is modified to
\bea
N=61.1 +\Delta N -\ln \bigg(\frac{V^{1/4}_0}{H_k} \bigg) -\frac{1}{12} \ln \bigg(\frac{g_{\rm RH}}{106.75} \bigg)
\eea
where the correction to the number of efoldings due to the non-instantaneous reheating is given by
\bea
\Delta N= \frac{1}{12} \bigg(\frac{3w-1}{w+1} \bigg) \ln\bigg(\frac{45 V_0}{\pi^2 g_{\rm RH} T^4_{\rm RH}} \bigg). \label{DN}
\eea
Here, $H_k$ is the Hubble parameter evaluated at the horizon exit for the Planck pivot scale, $k=0.05\,{\rm Mpc}^{-1}$, and $w$ is the averaged equation of state during reheating.

We can introduce the $Z_2$ invariant renormalizable couplings of the waterfall fields to the SM Higgs $H$, as follows,
\bea
{\cal L}_{H} = - \kappa_1 (\chi^2_1+\chi^2_2)|H|^2 -\kappa_2 \chi_1\chi_2 |H|^2.
\eea
Then, taking the waterfall field to $\chi_1= v_\chi+\chi_c(t)$ with $\chi_c(t)$ being the waterfall condensate, we get the decay rate of the waterfall condensate  into Higgs fields as
\bea
\Gamma_{\chi_1}&=& \frac{\kappa_1^2 v^2_\chi }{2\pi m_1} \sqrt{1-\frac{4m^2_H}{m^2_1}}. \label{decay}
\eea
On the other hand, we note that the two-body decay mode of the waterfall condensate, $\chi_c\to \chi_2\chi_2$, is kinematically closed.
As a consequence, from eqs.~(\ref{RH}) with eq.~(\ref{decay}) and $m^2_1=2\lambda_\chi v^2_\chi$, we obtain the reheating temperature approximately by 
\bea
T_{\rm RH}\simeq \bigg(\frac{90}{\pi^2 g_{\rm RH}} \bigg)^{1/4} \bigg( \frac{\kappa^2_1}{4\pi \lambda_\chi}\bigg)^{1/2} \sqrt{M_P m_1}.
\eea
Therefore, for $\Gamma_{\chi_1}\ll  H_I\sim m_1$, namely,  $\kappa^2_1\ll 4\pi \lambda_\chi$, and taking $H_I\lesssim 1.6\times 10^{10}\,{\rm GeV}$ for $f\lesssim 10^{16}\,{\rm GeV}$, we get the reheating temperature as $T_{\rm RH}\ll  10^{14}\, {\rm GeV}$.  Taking $w=0$ and $g_{\rm RH}=106.75$ in eq.~(\ref{DN}), we obtain the number of efoldings  as
\bea
N=51.3+\frac{1}{3}\ln\bigg(\frac{H_I}{1.6\times 10^{10}\,{\rm GeV}}\bigg)+\frac{1}{3}\ln\bigg(\frac{T_{\rm RH}}{10^{14}\,{\rm GeV}}\bigg).
\eea
As a result, there is a wide range of the parameter space for a successful inflation, and a sufficiently large reheating temperature is achieved due to the decay of the waterfall field.

\section{Microscopic realizations}

In this section, we present a concrete microscopic model for the pNGB inflaton and the effective waterfall field couplings. 
For this purpose, we consider the waterfall couplings to light dark quarks $d, d^c$ and heavy dark quarks $u_i,u^c_i (i=1,2)$ in a dark QCD \footnote{A similar model for dark QCD was considered for the single waterfall coupling without a $Z_2$ symmetry in Ref.~\cite{jeong}.}, with the Lagrangian, 
\bea
{\cal L}_{dQCD}=-m_u u_1 u^c_1-m_u u_2 u^c_2  -y \Phi_1 u^c_1 d-y' \Phi_1 u_1 d^c -iy \Phi_2 u^c_2 d-iy' \Phi^*_2 u_2 d^c +{\rm h.c.}
 \label{micromodel}
\eea
where  we imposed a $Z_2$ symmetry by $\Phi_1\to i\Phi_2$ and $\Phi_2\to-i\Phi_1$, and $y,y'$ are the Yukawa couplings taken to be real. We note that the waterfall fields, $\chi_1, \chi_2$, in our model, are  regarded as the real parts of the complex scalar fields, $\Phi_1, \Phi_2$, carrying the same dark PQ charges.

For $m_d\ll \Lambda_h\ll m_u$ with $\Lambda_h$ being the dark QCD scale, after integrating out $u_i, u^c_i$ and plugging $\Phi_{1,2}=\frac{1}{\sqrt{2}} \chi_{1,2} \, e^{i\phi/(4f)}$ into eq.~(\ref{micromodel}), we obtain the effective Yukawa couplings for $d,d^c$, as follows,
\bea
{\cal L}_{\rm dQCD, eff} &=&-\frac{yy'}{m_u} (\Phi^2_1-\Phi^2_2)  \, d d^c +{\rm h.c.} \nonumber \\
&=&-\frac{yy'}{2m_u}(\chi^2_1-\chi^2_2)  \, e^{i\phi/(2f)}\, d d^c +{\rm h.c.}. \label{dquarkmass}
\eea
Then, after the dark QCD condensation, integrating out dark mesons and making a shift by $\phi/(2f)\to \phi/(2f)+\pi$, we get the effective waterfall field couplings by
\bea
{\cal L}_{\rm eff}=\frac{1}{2}\mu^2 (\chi^2_1-\chi^2_2) \sin\Big(\frac{\phi}{2f} \Big) \label{effwater}
\eea
with
\bea
\mu^2 =\frac{|y y'|}{m_u}\, \Lambda^3_h. \label{micro}
\eea
Thus, the $Z_2$ symmetry in the dark QCD remains unbroken in the effective interactions generated after the dark QCD condensation.
On the other hand, we remark that the inflaton potential in eq.~(\ref{inflaton}) can be also originated from an extra dark QCD condensation at a scale $\Lambda$, respecting the $Z_2$ symmetry with $\phi\to -\phi$. 

We remark that after the inflation ends, the effective dark quark mass for $d,d^c$ in eq.~(\ref{dquarkmass}) depends on the VEVs of the inflaton and the waterfall fields, as follows,
\bea
m_{d,{\rm eff}} = \frac{\mu^2v^2_\chi}{\Lambda^3_h}.  \label{dquarkeff}
\eea
Therefore, from $m_{d,{\rm eff}}\ll \Lambda_h \ll m_u$ with eqs.~(\ref{micro}) and (\ref{dquarkeff}), we get $\mu^2 v^2_\chi \ll \Lambda^4_h$ and $\mu^2\ll |yy'| \Lambda^2_h$, for which the effective description for the waterfall field couplings in eq.~(\ref{effwater}) is valid.
Consequently, from $\Lambda^4\ll \mu^2 v^2_\chi\sim 4V_0 \ll \Lambda^4_h$, we need a hierarchy of the condensation scales by $\Lambda_h\gg \Lambda$.

We also remark that the rest of the scalar potential for the waterfall sector in eq.~(\ref{full}) is originated from the $Z_4$-invariant potential for $\Phi_1, \Phi_2$, as follows,
\bea
\Delta V_W=m^2_\chi( |\Phi_1|^2+|\Phi_2|^2)+ \lambda_\chi (|\Phi_1|^4+|\Phi_2|^4)+2{\bar\lambda}_\chi  |\Phi_1|^2 |\Phi_2|^2.
\eea
We note that $i(\Phi_1\Phi^\dagger_2-\Phi_2 \Phi^\dagger_1)$ is also $Z_2$-invariant, but it does not contribute to the scalar potential.
Therefore, the scalar potential for the waterfall sector with $\alpha=0$ in eq.~(\ref{full}) can be realized in the above microscopic model for dark QCD.

\section{Conclusions}

We have presented a successful model for natural inflation with twin waterfall fields where the $Z_2$ symmetry protects the inflaton potential from quantum corrections of waterfall fields during inflation. 
For the waterfall transition to end inflation,  we had to  choose the waterfall field couplings appropriately, but there is neither trans-Planckian axion decay constant or fine-tuning of the initial condition for the inflation. As far as the waterfall masses are parametrically smaller than the QCD-like condensation scale responsible for the inflaton potential,  we can leave the waterfall fields decoupled safely during inflation, while keeping the inflationary predictions under control.

In the post-inflationary regime, we obtained the VEVs of the waterfall fields in the $Z_2$ breaking vacuum by $v_\chi\sim \sqrt{V_0}/\mu$ where the inflation scale is $V_0$ and  the waterfall couplings are given by $\mu$. Thus, the inflaton receives a large mass contribution from the waterfall field couplings, while the physical masses for the waterfall fields are of similar order as those during inflation. 
Therefore, we can relate the waterfall couplings for inflation to the physical parameters for post-inflation at low energy.

We also introduced $Z_2$-invariant couplings between the waterfall fields and the SM Higgs for reheating and showed that the perturbative decay of the waterfall field is dominant for determining the reheating temperature. We found that the waterfall couplings to the SM Higgs are crucial to determine the reheating temperature as well as the number of efoldings more precisely.
In a concrete QCD-like microscopic model for inflation, we showed that the twin complex scalar fields with the Yukawa couplings to light and heavy quarks in the dark QCD can unify the inflaton and the waterfall fields with a desired form of the scalar potential respecting the $Z_2$ symmetry.

\section*{Acknowledgments}

The work is supported in part by Basic Science Research Program through the National
Research Foundation of Korea (NRF) funded by the Ministry of Education, Science and
Technology (NRF-2022R1A2C2003567 and NRF-2021R1A4A2001897).

\def\theequation{A.\arabic{equation}}

\setcounter{equation}{0}

\vskip0.8cm
\noindent
{\Large \bf Appendix A: The general vacuum for waterfall fields}

For a mixing mass for waterfall fields ($\alpha\neq 0$), there is a stable minimum of the potential at $\langle\phi\rangle=v_\phi$ and the VEVs for both waterfall fields, namely,  $\langle\chi_1\rangle=v_1\equiv v_\chi \cos\beta$ and  $\langle\chi_2\rangle=v_2\equiv v_\chi \sin\beta$, with
\bea
v_\phi &=& \pi f, \\
 v_\chi&=& \sqrt{\frac{1}{\lambda_\chi} \Big(\frac{\mu^2}{\cos2\beta}-m^2_\chi \Big)}, \label{chivevg}
\eea
and
\bea
(\mu^2+m^2_\chi+\lambda_\chi  v^2_\chi)\tan^3\beta +(\mu^2+m^2_\chi+{\bar\lambda}_\chi v^2_\chi) \tan\beta -\alpha^2(1+\tan^2\beta)=0. \label{tanbeta}
\eea
Thus, the real solution for $\tan\beta$ to eq.~(\ref{tanbeta}) is given by
\bea
\tan\beta= \frac{\alpha^2}{3\mu^2} +(R+\sqrt{R^2+Q^3})^{1/3} + (R-\sqrt{R^2+Q^3})^{1/3},
\eea
 for $R^2+Q^3>0$,
with
\bea
R&=& \frac{\alpha^2}{24\mu^4} \Big(18\mu^2-9(m^2_\chi+{\bar\lambda}_\chi v^2_\chi)+\frac{2\alpha^4}{\mu^2} \Big), \\
Q&=&\frac{1}{9\mu^2} \Big( 3(\mu^2 + m^2_\chi +{\bar\lambda}_\chi v^2_\chi) -\frac{\alpha^4}{\mu^2}\Big).
\eea
Here, we find that for $\alpha=0$, the stable minimum exists only for $\tan\beta=0$, namely, $\langle\chi_1\rangle=v_\chi$ and $\langle\chi_2\rangle=0$, recovering the results in Section 4.1. In general, from eq.~(\ref{chivevg}), the $Z_2$ symmetric vacuum with $v_1=v_2$ does not exist, so the $Z_2$ symmetry is necessarily broken in the vacuum. 

We find that the cosmological constant in the true vacuum  is fine-tuned to the observed value by
\bea
V_{\rm eff}(\chi_1=v_\chi,\chi_2=0) = V_0-\Lambda^4- \frac{1}{4}\lambda_\chi (v^4_1+v^4_2)-\frac{1}{2} {\bar\lambda}_\chi v^2_1 v^2_2 \simeq 0, \label{zeroccg}
\eea
constraining the parameters of the waterfall fields.

Expanding around the VEV by $\phi=v_\phi+a$  we obtain the inflaton mass as
\bea
m^2_{a} = \frac{1}{f^2} \Big(\Lambda^4+\frac{1}{8} \mu^2 v^2_\chi\cos2\beta\Big), \label{inflatonmassg} 
\eea
thus the inflaton receives a mass contribution from the waterfall field couplings. 
On the other hand, for $\chi_{1,2}=v_{1,2}+{\tilde\chi}_{1,2}$, the squared mass matrix for the waterfall fields is given by
\bea
{\cal M}^2_\chi=\left(\begin{array}{cc} 2\lambda_\chi v^2_\chi\cos^2\beta+\alpha^2\tan\beta & 2{\bar\lambda}_\chi v^2_\chi \sin\beta\cos\beta-\alpha^2 \vspace{0.2cm} \\   2{\bar\lambda}_\chi  v^2_\chi \sin\beta\cos\beta-\alpha^2  & 2\lambda_\chi v^2_\chi \sin^2\beta+\alpha^2\cot\beta  \end{array} \right).
\eea
Then, the mass eigenvalues for the waterfall fields are
\bea
m^2_{1,2} &=& \frac{1}{2} \bigg( 2\lambda_\chi v^2_\chi + \alpha^2 (\tan\beta+\cot\beta) \nonumber \\
&&\mp \sqrt{(2\lambda_\chi v^2_\chi \cos2\beta+\alpha^2(\tan\beta-\cot\beta))^2+4({\bar\lambda}_\chi v^2_\chi\sin2\beta-\alpha^2)^2}  \bigg),
\eea
and the mixing angle between the waterfall fields is
\bea
\sin2\theta = \frac{2({\bar\lambda}_\chi v^2_\chi \sin2\beta-\alpha^2)}{m^2_2-m^2_1}.
\eea
Here, a stable minimum for the waterfall fields exists, provided that
\bea
\lambda^2_\chi-{\bar\lambda}^2_\chi+\frac{\alpha^2}{2v^2_\chi\sin\beta\cos\beta}\Big(\lambda_\chi(\cos^4\beta+\sin^4\beta)+2{\bar\lambda}_\chi \sin^2\beta\cos^2\beta \Big)>0. \label{stableming}
\eea
We also note that the vacuum stability requires $\lambda_\chi>0$ and $\lambda_\chi+{\bar\lambda}_\chi>0$ for ${\bar\lambda}_\chi<0$. Therefore, as far as the vacuum stability conditions are satisfied, eq.~(\ref{stableming}) is automatically satisfied.

As a result, the leading interaction terms between the inflaton and the mass eigenstates of the waterfall fields, $ {\hat\chi}_{1,2}$, are as follows,
\bea
{\cal L}_{\rm int} &=& \frac{\mu^2}{8f^2}\,v_\chi a^2\Big((\cos(\beta-\theta)\, {\hat\chi}_1-\sin(\beta-\theta)\, {\hat\chi}_2\Big) \nonumber \\
&&+ \frac{\mu^2}{16f^2} a^2 \Big(\cos2\theta({\hat\chi}^2_1-{\hat\chi}^2_2)+\sin2\theta\,{\hat\chi}_1{\hat\chi}_2\Big) +\cdots. \label{inflaton-intg}
\eea
Thus, as in the case with $\alpha=0$ in the text, the quadratic divergences in the radiative inflaton mass are cancelled between the waterfall fields with the $Z_2$ symmetry, but the radiative inflaton mass is log-divergent due to the cubic interactions in the first line in eq.~(\ref{inflaton-intg}).

\end{document}